\documentstyle[12pt]{article}
\newcommand{\ra}{\rangle}
\newcommand{\la}{\langle}
\newcommand{\im}{{\rm i}}
\newcommand{\be}{\begin{equation}}
\newcommand{\ee}{\end{equation}}

\begin{document}

\begin{center}
\medskip

\textbf{Coupled Cluster Treatment of the XY model}

\bigskip

D.J.J. Farnell, S.E. Kr\"uger$^*$ and J.B. Parkinson

\bigskip

\textit{Department of Physics, UMIST, P.O.Box 88,
Manchester M60 1QD.}
\end{center}

\bigskip

\medskip\ 
\noindent \underline{Abstract} We study quantum spin systems in the 1D, 2D square and 3D cubic lattices with nearest-neighbour XY exchange. We use the coupled-cluster method (CCM) to calculate the ground-state energy, the $T=0$ sublattice magnetisation and the excited state energies, all as functions of the anisotropy parameter $\gamma$. We consider $S=1/2$ in detail and give some results for higher $S$. In 1D these results are compared with the exact $S=1/2$ results and in 2D with Monte-Carlo and series expansions. We obtain critical points close to the expected value $\gamma=0$ and our extrapolated LSUBn results for the ground-state energy are well converged for all $\gamma$ except very close to the critical point. 

\bigskip

\noindent  * Permanent address: Institut f\"ur Theoretische Physik, Universit\"at Magdeburg, P.O. Box 4120, D-39016 Magdeburg, Germany


\medskip
\noindent \underline{Short Title} CCM for the XY model.

\newpage\ 

\medskip
\noindent \textbf{1. Introduction and CCM formalism}

In this paper we consider the $T=0$ properties of the quantum spin system known as the XY-model, described by the Hamiltonian

\begin{equation} 
H=\frac{1}{2}\sum_{l,p}[(1+\gamma)s_l^xs_{l+p}^x+(1-\gamma)s_l^ys_{l+p}^y] \qquad \mbox{in the regime }0\le \gamma \le 1 
\end{equation}

\noindent where index $l$ runs over all $N$ lattice sites with periodic boundary conditions, and $p$ over the $z$ nearest-neighbour sites. 

For $s=1/2$ and 1D this model was solved exactly by Lieb, Schultz and Mattis (1961) and its properties have been studied by many authors (see Niemeyer, 1967, and Barouch \textit{et al.}, 1971, for example). For higher spin in 1D or in 2D (square) and 3D (simple cubic) useful results have been obtained using spin-wave theory (Zheng \textit{et al.}, 1991), Monte-Carlo methods (Ding, 1992, Zhang and Runge, 1992), series expansions (Hamer \textit{et al.}, 1991) and, for $\gamma=0$, finite size extrapolations (Betts \textit{et al.}, 1996).

In a recent paper (Bishop, Farnell and Parkinson, 1996, referred to as I), the coupled-cluster method (CCM) was applied to the XXZ model in the $|\Delta|<1$ regime. It was found that good results could be obtained by using a \textit{planar} model state in which the spins are aligned in the $xy-$plane, as in the classical ground state, rather than along the $z-$axis. Here we shall use a similar model state for (1), again motivated by the classical ground state.

For a description of the CCM applied to spin systems see Bishop \textit{et al. }(1991) and also the references given in I. To calculate the ground state wave function $|\Psi\ra$ of a spin system we start with a \underline{model state} $|\Phi\ra$ and a \underline{correlation operator} $S$ such that

\[
|\Psi\ra = e^S|\Phi\ra
\]

For the Hamiltonian (1) we expect that in the ground state the spins are aligned in the $xy-$plane. We choose $|\Phi\ra$ to be a N\'eel state with spins aligned parallel and antiparallel to the $x-$axis. In 1D this has the form   

\[
|\Phi\ra = |\ldots \quad \longleftarrow \quad \longrightarrow \quad \longleftarrow \quad \longrightarrow \quad \longleftarrow \quad \longrightarrow \quad \longleftarrow \quad \longrightarrow \quad \longleftarrow \quad \longrightarrow \quad \ldots \ra.
\]

It is useful to introduce \underline{local axes} such that each spin in $|\Phi\ra$ is pointing in the negative $z-$direction, by means of the following transformation:

\[ 
s^x\rightarrow -s^z, s^y\rightarrow s^y, s^z\rightarrow s^x \quad\mbox{left-pointing spins} 
\]
\[ 
s^x\rightarrow s^z,  s^y\rightarrow s^y, s^z\rightarrow -s^x \quad\mbox{right-pointing spins.} 
\]

Thus (1) becomes (with $s^{\pm}=s^x\pm \im s^y$)

\be 
H=\frac{1}{2}\sum_{l,p}[As_l^zs_{l+p}^z+B(s_l^+s_{l+p}^++s_l^-s_{l+p}^-)+C(s_l^+s_{l+p}^-+s_l^-s_{l+p}^+)]
\ee

with
\[ 
A\equiv -(1+\gamma), \quad B\equiv -\frac{1}{4}(1-\gamma), \quad C=-B.
\]

For the correlation operator $S$ we choose a linear combination of creation operators relative to $|\Phi\ra$, a creation operator being any combination of spin raising operators ($s^{+}$ in the local axes). Because of the form of (2) the total number of spin flips in each creation operator must be even.

The simplest possible choice for $S$ is to flip two spins, known as the SUB2 approximation scheme: 

\be 
S=\sum_{l=1}^N\left(\frac{1}{2}\sum_rb_rs_l^+s_{l+r}^+\right) 
\ee

\noindent where $r$ runs over all distinct lattice vectors ($r\ne0$ for $s=1/2$).

The full SUB4 scheme involves 4-flip configurations as well 2 flips and is too complicated to handle in general. However the most important extra term is the one with 4 flips on adjacent sites. Including this term gives the SUB2+LSUB4 scheme which we have applied only in 1D:

\be 
S=\sum_{l=1}^N\left(\frac{1}{2}\sum_rb_rs_l^+s_{l+r}^++g_4s_l^+s_{l+1}^+s_{l+2}^+s_{l+3}^+\right) 
\ee

A third approximation scheme is to include in $S$ all possible combinations   of spin flips within a region of size $n$, known as the LSUBn scheme. This is particularly useful for numerical extrapolation as a function of $n$, and will be discussed in detail in section 5.

From the Schr\"odinger equation $H|\Psi\ra=E|\Psi\ra$ we obtain the equation for the ground state energy:

\be 
E=\la\Phi| e^{-S}He^S|\Phi\ra=\frac{1}{2}zN\left(\frac{1}{4}A+b_1B\right) 
\ee

\noindent This equation is exact whatever approximations are made for $S$. 

To determine the coefficients $b_r$ and $g_4$ in the SUB2+LSUB4 scheme we operate on the Schr\"odinger equation with $\exp(-S)$ then one of the destruction operators and then by $\la\Phi|$:

\newpage
\[
\la\Phi|s_l^-s_{l+r}^-e^{-S}He^S|\Phi\ra=\displaystyle\sum_p\Bigl[B\sum_{r'}
b_{r'}b_{r-r'+p} -(A+4Bb_1)b_r
\]
\be
+2Cb_{r-p}+\bigl(B(2b_1^2+2g_4+1)+Ab_1\bigr)\delta_{p,r}+Bg_4\delta_{3p,r}\Bigr]=0 
\ee

\[
\la\Phi|s_l^-s_{l+1}^-s_{l+2}^-s_{l+3}^-e^{-S}He^S|\Phi\ra=A(b_1^2+3b_2^2
+2b_1b_3)-4B(b_1b_2b_4+b_1b_3^2+b_2^2b_3)
\]
\be
-4C(2b_1b_2+b_2b_3)+ g_4\bigl[B(2b_5-2b_3-8b_1)-A\bigr]=0 
\ee

\noindent The corresponding equations for the SUB2 scheme are obtained by 
setting $g_4=0$ everywhere in the first of these and ignoring the second.

These coupled \underline{non-linear} equations are solved by first Fourier 
transforming Eq.(6) and then solving the resulting equations and Eq.(10) self-consistently. For dimension $d$ we obtain:

\[ 
\Gamma(q)\equiv\sum_re^{\im rq}b_r, \quad b_r=\int_{-\pi}^{\pi}d^dq(2\pi)^{-d}e^{-\im rq}\Gamma(q),\quad \gamma(q)=\frac{1}{z}\sum_pe^{\im pq}  
\]
\[ 
b_1=\int_{-\pi}^{\pi}d^dq(2\pi)^{-d}\gamma(q)\Gamma(q), \quad X_1\equiv\sum_rb_rb_{r+p}=\int_{-\pi}^{\pi}d^dq(2\pi)^{-d}\gamma(q)\Gamma^2(q) \]
leading to
\[ 
a\Gamma^2(q)+b\Gamma(q)+c=0, 
\]
where
\[ 
a\equiv B\gamma(q) \mbox{ , }\quad b\equiv -A-4Bb_1+2C\gamma(q), 
\]
\[
c\equiv [B(2b_1^2+2g_4+1)+Ab_1)]\gamma(q)+Bg_4\gamma(3q)-BX_1-2Cb_1 
\]
with the usual solution
\[
\Gamma(q)=\frac{-b+\sqrt{b^2-4ac}}{2a}.
\]

\noindent The equations can now be solved numerically in a self-consistent way.

Results for the ground state energy using the SUB2 and SUB2+LSUB4 approximation schemes are shown in Figures 1 and 3, and Tables 1 and 2 for 1D and 2D. The LSUBn results are discussed in section 5.

A notable feature of the CCM is the existence of terminating points as a function of $\gamma$. These are believed to correspond to the actual $T=0$ phase changes, known to be at $\gamma=0$ in 1D and believed also to be at $\gamma=0$ on symmetry grounds for 2D and 3D. In 1D terminating points only occur if correlations of infinite range are explicitly included in $S$ and occur at $\gamma=-0.10789$ in SUB2 and at $\gamma=-0.09605$ in SUB2+LSUB4. In 2D there is a terminating point at $\gamma=-0.03033$ in SUB2. These are reasonably close to $\gamma=0$ considering the simple nature of these approximations.

In 2D and 3D terminating points can also occur within the LSUBn scheme as described in section 5.

\bigskip
\bigskip
\noindent \textbf{2. In-plane Sublattice Magnetisation}

In the CCM the bra ground state is not in general the Hermitian conjugate of 
the ket state. Instead we introduce a new operator $\tilde S$ such that 

\[
\la \tilde \Psi| = \la \Phi|\tilde S \exp(-S)\mbox{  }.
\]

 The SUB2+LSUB4 approximation for $\tilde S$ is

\be 
\tilde S=1+\sum_{l=1}^N\left(\frac{1}{2}\sum_r\tilde b_rs_l^-s_{l+r}^-
+\tilde g_4s_l^-s_{l+1}^-s_{l+2}^-s_{l+3}^-\right)
\ee

\noindent where $r$ runs over all distinct lattice vectors (with $r\ne0$ for $s=1/2$).

The bra-state equations are found variationally by taking the partial derivatives of  

\[
\bar H=\la \tilde\Psi|H|\Psi\ra
\]

\noindent with respect to the ket-state coefficients. By CCM theory (Bishop \textit{et al.} 1991) these derivatives must be equal to $0$. Hence we obtain two bra state equations:

\begin{eqnarray}
 \displaystyle\frac{\partial \bar H}{\partial b_r}&=&N\displaystyle\sum_p\Bigl[2B\sum_{r'}\tilde b_{r'}b_{r'-r+p}
  -(A+4Bb_1)\tilde b_r+2C\tilde b_{r-p} \nonumber \\
  &&+\bigl(B+(A+4Bb_1)\tilde b_1-4B\displaystyle\sum_{r'}\tilde b_{r'}b_{r'}\bigr)\delta_{p,r} \nonumber \\
  &&+\tilde g_4/2\bigl\{[4A(b_1+b_3)-8B(b_2b_4+b_3^2)-16Bg_4-16Cb_2]\delta_{p,r} \nonumber \\
  &&+ [12Ab_2-8B(b_1b_4+2b_2b_3)-8C(2b_1+b_3)]\delta_{2p,r} \nonumber \\
  &&+ [4Ab_1-8B(2b_1b_3+b_2^2)-4Bg_4-8Cb_2]\delta_{3p,r} \nonumber \\
  &&-8Bb_1b_2\delta_{4p,r}+4Bg_4\delta_{5p,r} \bigr\} \Bigr]=0 
\end{eqnarray} 

\be 
\frac{\partial \bar H}{\partial g_4}=N\left[B(2\tilde b_1+\tilde b_3)
+\tilde g_4(2B(b_5-b_3-4b_1)-A)\right]=0 
\ee

\noindent Again we perform a Fourier transform on Eq.(9) and the resulting equations and Eq.(10) may be solved self-consistently in order to obtain the bra-state correlation coefficients.

Finally the results are used to calculate the magnetisation using the 
formula for SUB2:

\[ 
M=-2\la\tilde\Psi|s_l^z|\Psi\ra=1-2\sum_r\tilde b_rb_r 
\]

\noindent and for SUB2+LSUB4:

\[ 
M=-2\la\tilde\Psi|s_l^z|\Psi\ra=1-2\sum_r\tilde b_rb_r-8g_4\tilde g_4 
\]

\bigskip
\bigskip
\noindent \textbf{3. Excitations}

A similar method can be used for the excited state energies, introducing the 
operator 
\[ 
X_1=\sum_i{\cal X}_i s_i^+, \quad \mbox{\textit{i} belongs to one sublattice only}
\]

leading to

\be 
\la\Phi| s_l^- e^{-S}[H,X_1]-e^S|\Phi\ra=-\frac{1}{2}z(A+4Bb_1){\cal X}_l+B\sum_{r,p}b_r{\cal X}_{l+r+p}=\varepsilon_l{\cal X}_l 
\ee

and hence, via Fourier transform

\be 
\Rightarrow \varepsilon(q)=-\frac{1}{2}z(A+4Bb_1)+Bz\gamma(q)\Gamma(q) 
\ee

\bigskip
\bigskip
\noindent \textbf{4. General spin $s$, (SUB2 only)}

We have also considered the general case of $s\ge 1/2$ within the SUB2 
approximation scheme. The main features are as follows.

The correlation operators $S$ and $\tilde S$ are the same as before 
(without $g_4$). The ket state equations become:

\begin{eqnarray}
\la\Phi| s_l^-s_{l+r}^- e^{-S}He^S|\Phi\ra&=&4s^2\displaystyle\sum_p\Bigl[4s^2B\sum_{r'}b_{r'}b_{r-r'+p}
  -2s(A+4Bb_1)b_r+4sCb_{r-p} \nonumber \\
  &&+(B(2b_1^2+1)+Ab_1)\delta_{p,r}\Bigr]=0 
\end{eqnarray} 

and the energy is 
\be 
\la\Phi| e^{-S}He^S|\Phi\ra=2s^2zN\left(\frac{1}{4}A+b_1B\right) 
\ee

Using these equations we find for the ground-state energy per spin of the $s=1$ system at the $\gamma=0$ point the value $-1.09179$. This compares with a numerical result from extrapolating rings with $N\le14$ of $-1.1157\pm 0.0003$. This is a very similar accuracy to that obtained using SUB2 for $s=\frac{1}{2}$ at the same point.

There are similar modifications to the bra state equations which become

\begin{eqnarray}
 \displaystyle\frac{\partial \bar H}{\partial b_r}&=&4s^2N\displaystyle\sum_p\Bigl[8s^2B\sum_{r'}\tilde b_{r'}b_{r'-r+p}
  -2s(A+4Bb_1)\tilde b_r+4sC\tilde b_{r-p} \nonumber \\
  &&+\bigl(B+(A+4Bb_1)\tilde b_1-8sB\displaystyle\sum_{r'}\tilde b_{r'}b_{r'}\bigr)\delta_{p,r}\Bigr]=0 
\end{eqnarray}

Finally the magnetisation is given by:

\be 
M=-\frac{1}{s}\la\tilde\Psi|s_l^z|\Psi\ra=1-4s\sum_r\tilde b_rb_r 
\ee

\bigskip

\noindent \textbf{5. The LSUB{\it n} Approximation}
\medskip

The LSUB{\em n} scheme contains all possible (connected and 
disconnected) terms in $S$ which are contained within a `locale' 
of size $n$. We use all possible connected 
configurations of $n$ spins to define this locale; in 1D we may 
see that this locale is simply a chain of length $n$. Disconnected 
and connected configurations of less than $n$ spins are then generated by 
successively removing sites from the original connected configurations of
$n$ spins, thus covering all possibilities. The lowest order LSUB{\em n} 
approximation scheme is the LSUB2 (i.e., SUB2-2) approximation in 
which only a single nearest-neighbour, two-body term is retained in $S$. 
We note that the Hamiltonian of Eq. (2) includes products of the 
spin operators which contain even numbers of these spin operators only. 
This means that the ground state contains only even numbers of spin 
flips with relation to the model state. We restrict the LSUB{\it n} 
approximation to include only those configurations which contain an 
even number of spin raising operators. A further restriction is that 
each {\it fundamental} configuration must be independent of all others 
under the symmetries of both the lattice and the Hamiltonian; 
we note that both lattice and the Hamiltonian have identical symmetries 
for the $XY$ model.

Tables \ref{tab1}, \ref{tab2}, and \ref{tab3}  show the numbers of 
fundamental configurations for given LSUB{\it n} approximation 
level, and we can see from these tables that the number of 
configurations grows very rapidly with $n$. Hence, for higher-order 
approximations we need to enumerate all possible configurations 
computationally, and we furthermore need to obtain and solve 
the CCM LSUB{\it n} equations computationally also. A full 
explanation of the computational method used here is given in 
Zeng \textit{et al.} (1997). It is now possible to obtain values for the 
ground-state energy and sublattice magnetisation for 
the LSUB{\it n} approximation scheme. Results for these
quantities are given in Figs. 1,2,3 and 4, and results at the isotropic point of $\gamma$=0 are given in Tables \ref{tab1}, \ref{tab2}, and \ref{tab3}. A simple extrapolation of the ground-state energy and sublattice magnetisation 
has also been carried out by plotting the ground-state energy against $1/n^2$ and the sublattice magnetisation against $1/n$, and then performing polynomial fits on this data. The extrapolated LSUB$\infty$ results obtained from this simple, `naive' approach are shown in Tables \ref{tab1}, \ref{tab2}, and \ref{tab3}. The results are clearly at least as good as obtained by series expansion. Results in 2D and 3D are especially valuable since no exact results are available.

Another consequence of this approximation scheme is that the second 
derivative of the ground-state energy is found to diverge for some
{\it critical} value of the anisotropy parameter, denoted $\gamma_c(n)$,
in 2D and 3D only. These points are related to phase transitions of 
the true ground state of the system (Zeng \textit{et al.}, 1997), and the results for given LSUB{\it n} approximation level are shown in Tables 
\ref{tab2}, and \ref{tab3}. We note that critical $\gamma_c(n)$ 
approaches $\gamma$=0, the point at which the true phase transition 
point is believed to be (in all dimensions), with increasing 
approximation level. Again, a simple extrapolation of the LSUB{\it n} 
critical points is attempted by plotting $\gamma_c(n)$ against 1/$n^2$, 
as in Bishop \textit{et al.} (1994), and the extrapolated results are also shown in Tables \ref{tab2} and \ref{tab3}. 

\bigskip
\noindent \textbf{6. Acknowledgements}

We have benefited from discussions with R.F. Bishop. D.J.J. Farnell acknowledges a research grant from the Engineering and Physical Sciences Research Council (EPSRC) of Great Britain. S.Kr\"uger is grateful to the Physics Department, UMIST, for the opportunity to visit to carry out this work.

\newpage\ 

\medskip

\noindent \textbf{References}

\medskip

\noindent Barouch E. and McCoy B.M., 1971, Phys. Rev. A, \textbf{3}, 786-804

\noindent Betts D.D, Matsui S., Vats N. and Stewart G.E., 1996, Can J Phys., \textbf{74}, 54-64

\noindent Bishop R.F., Parkinson J.B. and Yang Xian, 1991, Phys.Rev.B \textbf{ 44}, 9425-43

\noindent Bishop R.F., Farnell D.J.J. and Parkinson J.B., 1996, J.Phys.:CM \textbf{8}, 11153-65

\noindent Bishop R.F., Hale R.G. and Xian Y., 1994, Phys. Rev. Lett. \textbf{73}, 3157-60

\noindent Ding H-Q., 1992, Phys. Rev. B, \textbf{45}, 230-42

\noindent Hamer C.J., Oitmaa J. and Zheng W., 1991, Phys. Rev. B, \textbf{43}, 10789-96

\noindent Lieb E., Schultz T. and Mattis D., 1961, Annals of Phys. \textbf{16}, 407-66

\noindent McCoy B.M., 1968, Phys. Rev., \textbf{173}, 531-41

\noindent Niemeyer Th., 1967, Physica \textbf{36}, 377-419

\noindent Zeng C., Farnell D.J.J. and Bishop R.F., 1997, (to be published).

\noindent Zhang S. and Runge K.J., 1992, Phys. Rev. B, \textbf{45}, 1052-5

\noindent Zheng W., Oitmaa J. and Hamer C.J., 1991, Phys. Rev. B, \textbf{44}, 11869-81

\newpage\ 
\medskip

\noindent \textbf{Figure captions}

\medskip
\noindent \underline{Fig.1}

\noindent Results for the CCM ground-state energy of the one dimensional 
{\it XY} model. The terminating points of SUB2 and SUB2+LSUB4 schemes are indicated.

\medskip
\noindent \underline{Fig.2}

\noindent Results for the CCM ground-state sublattice magnetisation
of the one dimensional {\it XY} model.

\medskip
\noindent \underline{Fig.3}

\noindent Results for the CCM ground-state energy of the square lattice {\it XY} model. All the approximation schemes have terminating points except LSUB2. 

\medskip
\noindent \underline{Fig.4}

\noindent Results for the CCM ground-state sublattice magnetisation
of the square lattice {\it XY} model.

\newpage
\medskip

\begin{table} 
\caption{Ground-state energy and sublattice magnetisation for the 
one dimensional {\it XY} model at $\gamma=0$ compared to exact results
of McCoy (1968). N$_f$ indicates the number of fundamental
configurations for a given LSUB{\it n} approximation level.}

\makebox[1cm]{}

\begin{tabular}{|c|c|c|c|}  \hline\hline 
LSUB{\it n}	&N$_f$	&E$_g$/N	&M
\\ \hline\hline
LSUB2		&1	&$-$0.303813	&0.837286	
\\ \hline 
SUB2		&--	&$-$0.310377	&0.779517	
\\ \hline 
LSUB4		&4	&$-$0.314083	&0.722916	
\\ \hline 
LSUB6		&13	&$-$0.316301	&0.660064	
\\ \hline 
LSUB8		&43	&$-$0.317137	&0.617624	
\\ \hline 
LSUB10		&151	&$-$0.317542	&0.586067	
\\ \hline 
LSUB$\infty$	&--	&$-$0.31829	&--		
\\ \hline 
Exact		&--	&$-$0.318310	&0.0		
\\ \hline 
\end{tabular}
\label{tab1}
\end{table}

\begin{table} 
\caption{Ground-state energy and sublattice magnetisation for the 
square lattice {\it XY} model at $\gamma=0$ compared to 
series expansion calculations of Hamer, Oitmaa and Zheng [2]. 
N$_f$ indicates the number of fundamental configurations for a 
given LSUB{\it n} approximation level, and also shown are the 
critical values of $\gamma$ for the anisotropic model -- where the 
value in parentheses is the estimated error in the final decimal place
shown.}

\makebox[1cm]{}

\begin{tabular}{|c|c|c|c|c|}  \hline\hline 
LSUB{\it n}	&N$_f$	&E$_g$/N	&M		&$\gamma_c(n)$ 	
\\ \hline \hline
LSUB2		&1	&$-$0.540312	&0.949634	&--		
\\ \hline 
SUB2		&--	&$-$0.546325	&0.918953	&$-$0.030(1)
\\ \hline 
LSUB4		&10	&$-$0.547267	&0.915768	&$-$0.175(1) 
\\ \hline 
LSUB6		&131	&$-$0.548329	&0.901357	&$-$0.073(1)
\\ \hline 
LSUB8		&2793	&$-$0.548616	&0.893665	&$-$0.04(1)
\\ \hline 
LSUB$\infty$	&--	&$-$0.54892	&0.869		&0.00(1)
\\ \hline 
Series Expansion &--	&$-$0.5488 	&0.872 		&0.0
\\ \hline
\end{tabular}
\label{tab2}
\end{table}

\begin{table} 
\caption{Ground-state energy and sublattice magnetisation for the 
cubic lattice {\it XY} model at $\gamma=0$. N$_f$ indicates the 
number of fundamental configurations for a given LSUB{\it n} 
approximation level, and also shown are the critical values of 
$\gamma$ for the anisotropic model -- where the value in 
parentheses is the estimated error in the final decimal place
shown.} 

\makebox[1cm]{}

\begin{tabular}{|c|c|c|c|c|}  \hline\hline 
LSUB{\it n}	&N$_f$	&E$_g$/N	&M		&$\gamma_c(n)$ 	
\\ \hline \hline
LSUB2		&1	&$-$0.786866	&0.971488 	&--	
\\ \hline 
SUB2		&--	&$-$0.790901	&0.958282	&$-$0.01666(1)
\\ \hline 
LSUB4		&13	&$-$0.791224	&0.958648	&$-$0.172(1)
\\ \hline 
LSUB6		&327	&$-$0.791702	&0.954759	&$-$0.071(1)
\\ \hline 
LSUB$\infty$	&--	&$-$0.79201	&0.948		&0.01(1)
\\ \hline 
\end{tabular}
\label{tab3}
\end{table}

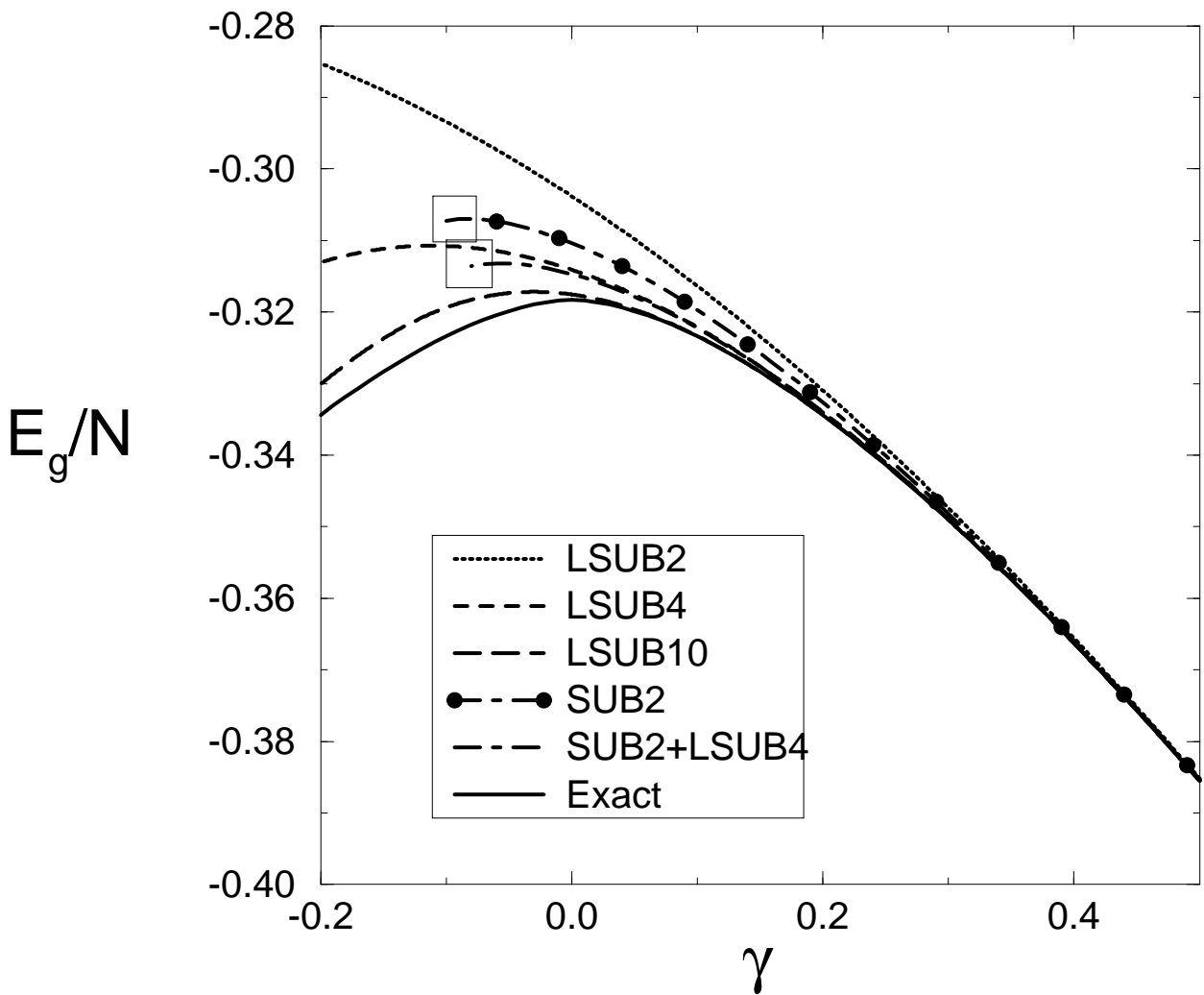
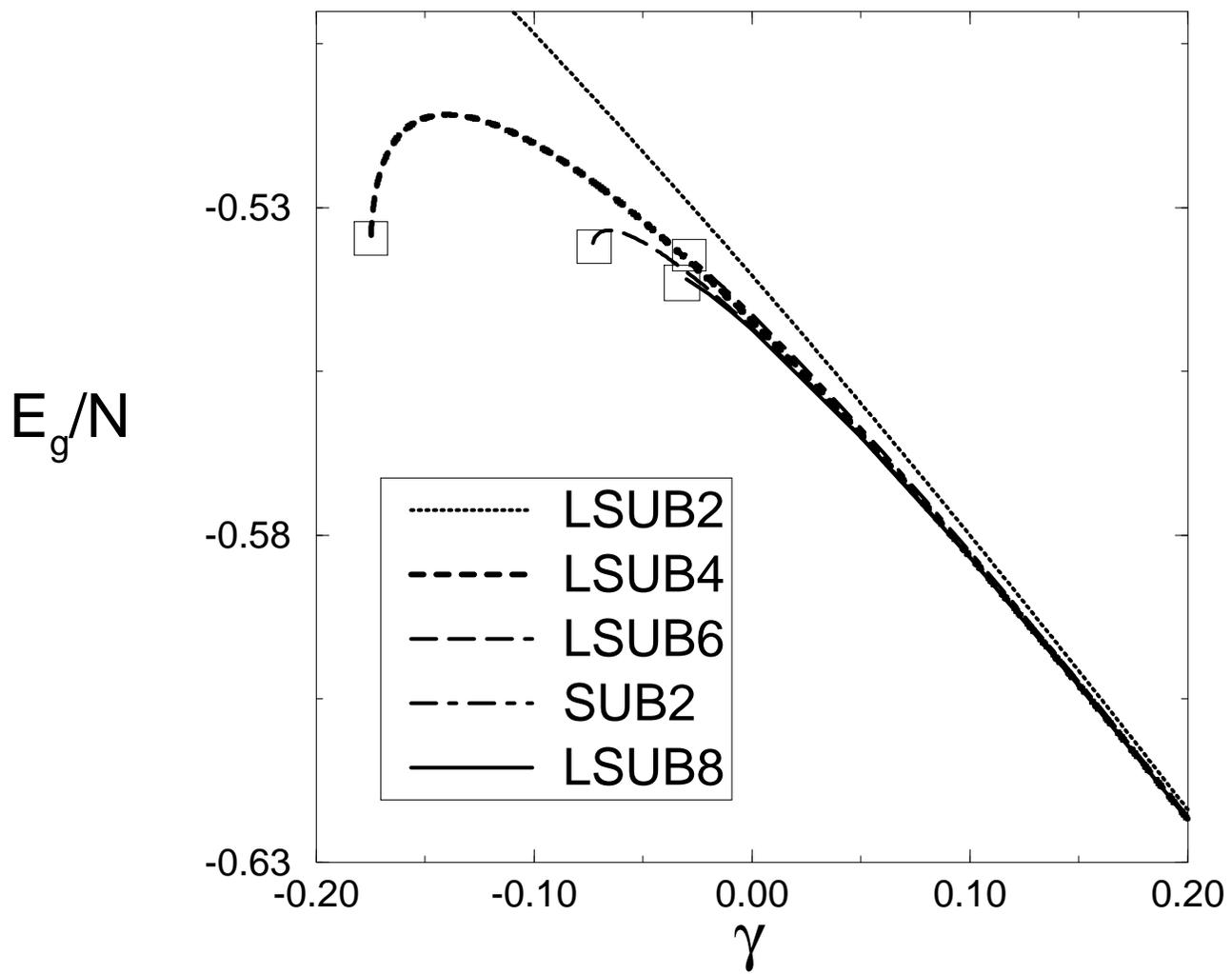
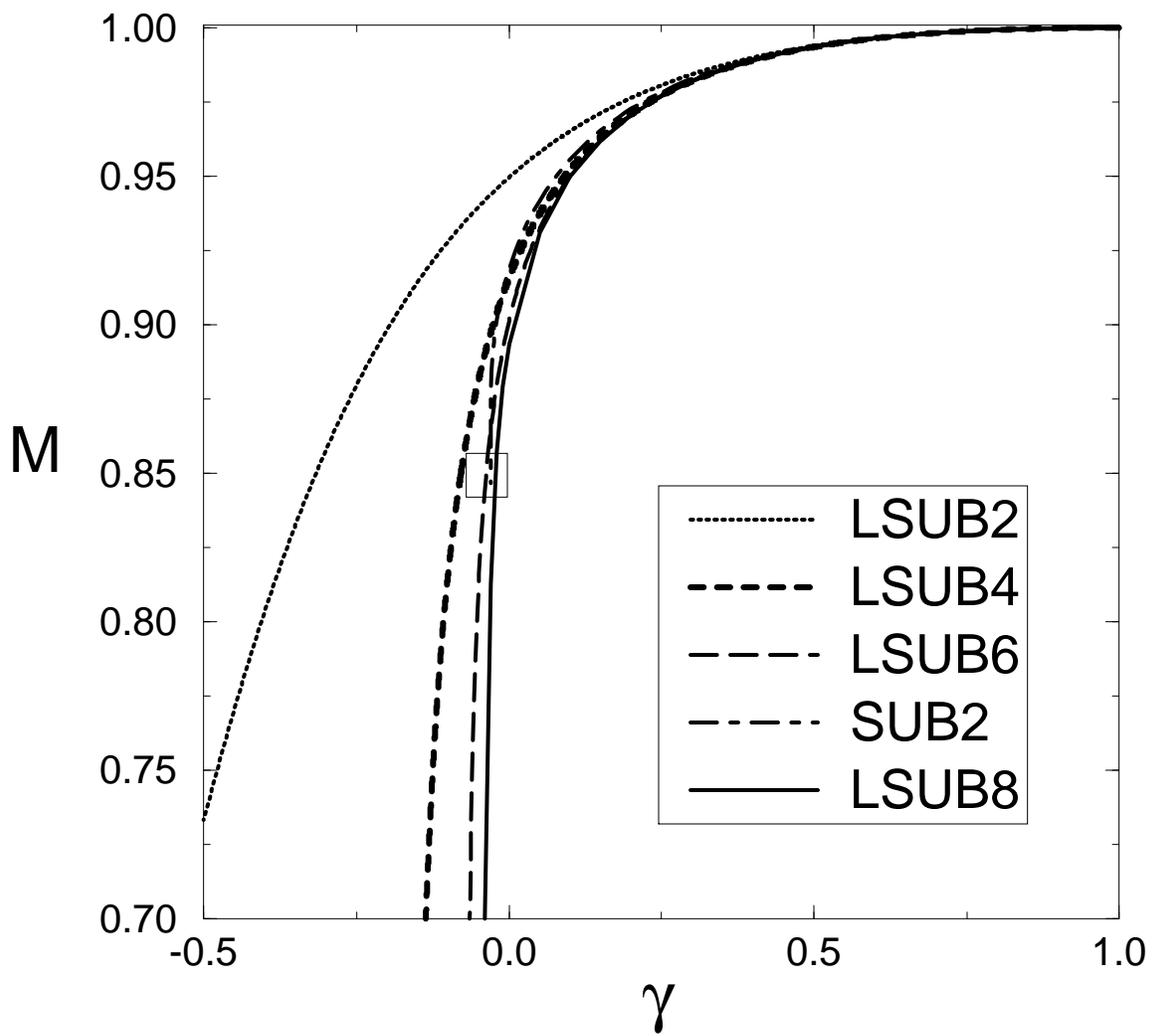
\end{document}